\documentclass[journal]{IEEEtran}

\ifCLASSINFOpdf
 
\else

\fi

\usepackage{graphicx}
\usepackage{cite}
\usepackage[T1]{fontenc}
\usepackage[utf8]{inputenc}
\usepackage{lmodern}
\usepackage{hyperref}

\begin{document}

\title{A Flipped Classroom Approach to Teaching Empirical Software Engineering}
%
%
%

\author{Lucas~Gren
\thanks{L. Gren is with the Department
of Computer Science and Engineering, Chalmers University of Technology and the University of Gothenburg, Gothenburg, Sweden, e-mail: lucas.gren@cse.gu.se}
}

\markboth{Preprint: IEEE Transactions on Education, Accepted December 8, 2019.}%
{Shell \MakeLowercase{\textit{et al.}}: Bare Demo of IEEEtran.cls for IEEE Journals}

\maketitle

\begin{abstract}
\\
\emph{Contribution}: A flipped classroom approach to teaching empirical software engineering increases student learning by providing more time for active learning in class.

\emph{Background}: 
There is a need for longitudinal studies of the flipped classroom approach in general. Although a few cross-sectional studies show that a flipped classroom approach can increase student learning by providing more time for other in-class activities, such as active learning, such studies are also rare in the context of teaching software engineering. 

\emph{Intended outcomes}: 
To assess the usefulness of a flipped classroom approach in teaching software engineering.

\emph{Application design}: 
The study was conducted at an international Master's program in Sweden, given in English, and partially replicated at a university in Africa.

\emph{Findings}: 
The results suggest that students' academic success, as measured by their exam grades, can be improved by introducing a flipped classroom to teach software engineering topics, but this may not extend to their subjective liking of the material, as measured by student evaluations. Furthermore, the effect of the change in teaching methodology was not replicated when changing the teaching team.

\end{abstract}
\begin{IEEEkeywords}
Active learning, blended learning, computer-based instruction, flipped classroom, multiculturalism, Master's students, software engineering
\end{IEEEkeywords}

%
\IEEEpeerreviewmaketitle

\section{Introduction}
In new information technology and active learning research \cite{freemanpnas,prince2004does}, many have called for the restructuring of engineering education to incorporate such results more systematically (see e.g., \cite{adams2011multiple}). One way of doing this is to flip the classroom. The idea of a flipped (or inverted) classroom does not, by itself, imply the use of modern technology. Instead, it is more related to active learning and can be defined as the idea of that ``events that have traditionally taken place inside the classroom now take place outside the classroom and vice versa'' \cite{lage2000inverting}. Therefore, providing students with a printed article, asking them to read it, and then discussing it with them in class is also a flipped classroom, and a technique that has been applied since long before computers were invented. On the other hand, \emph{blended learning} can be defined as ``the thoughtful integration of classroom face-to-face learning experiences with online learning experiences'' \cite{garrison2004blended}, which then implies the use of IT in combination with in-class activities. According to Garrison et al.  \cite{garrison2004blended}, blended learning is in line with the values of traditional higher education institutions and can ``enhance both the effectiveness and efficiency of and efficiency of meaningful learning experiences” \cite{garrison2004blended}.

\begin{figure}
\centerline{\includegraphics[scale=0.6]{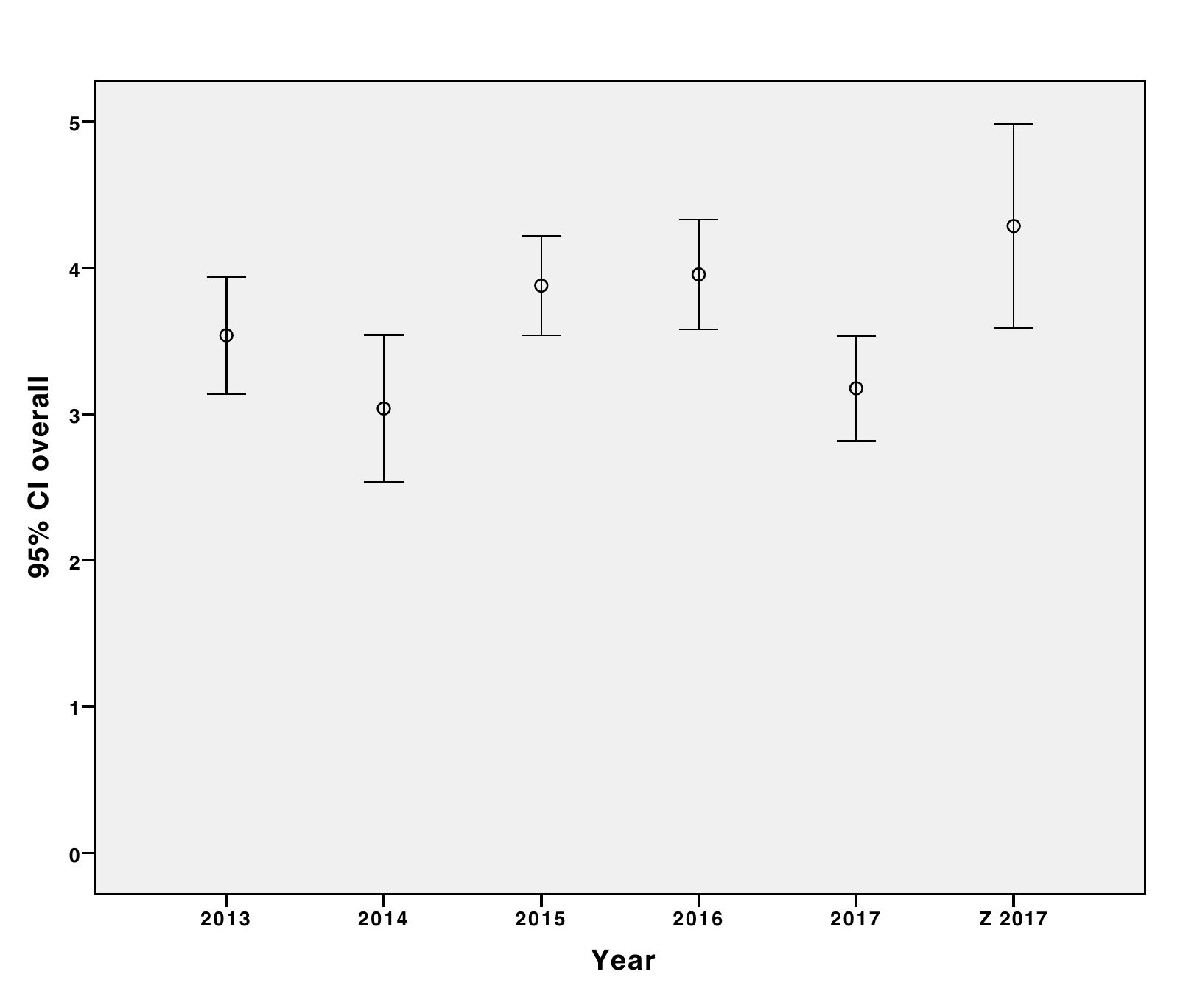}}
\caption{Students rating of their overall impression of the course across five years and from the University of Zambia. Note that the chart uses mean values in line with the university, but the statistical analysis below uses tests based on ranks since the data is not normally distributed.}
\label{overall2}
\end{figure}

In the present study, online and in-class components were always mixed, so the following definition of a flipped classroom is used: ``an educational technique that consists of two parts: interactive group learning activities inside the classroom, and direct computer-based individual instruction outside the classroom'' \cite{bishop2013flipped}. The background to this change in teaching methodology is very much connected to the academic life of the author of this present paper, who, when he started as a Ph.D.\ student in 2013, had 20\% teaching duties. In the second year of his studies, he was more involved in teaching a Master's-level course, Empirical Software Engineering, that included more statistical methods for empirical software engineering research to prepare students to apply quantitative methods to their thesis work. Until 2013, the course was mostly theoretical, which was not a good preparation for the students to use these methods when writing their theses. In the Fall of 2014, the author also took over some introductory statistics lectures from another teacher, and had lower student evaluation scores than in previous years (from a mean of 4 for the students' overall impression of the course to around 3). Fig.~\ref{overall2} shows the students' overall impression of the course from 2013 (when the author was first involved in some practical components of the course) until 2017. The 2014 student evaluation results raised the notion that there must be a better way than lecturing to teach hard topics to university students, and that the poor result was not only due to him being a novice lecturer. The fact that around a fifth of the students dropped out of the course in 2014 underlined the need for improvement of the course. While attending a pedagogical conference at Chalmers University of Technology\footnote{\url {https://www.chalmers.se/en/conference/KUL/} } in January 2015 he spoke with two people giving a short presentation on the flipped classroom, and they decided to try a flipped classroom approach to the course, if funding to change the course could be found, which it was shortly thereafter. The goal was to add to the body of knowledge in flipped classrooms for higher education, since such studies were scarce in 2015. In particular, using online components was seen as an easy change in methodology since software engineering students are very accustomed to IT. The greatest difficulty was that the teaching team needed a set of new skills. To increase the probability of a successful implementation, pedagogical flipped classroom experts were hired the first two years the course was flipped.

In the years since 2015, there has been increased interest in the flipped methodology, both in educational research and teaching practice. A quick overview of the results of those efforts is presented Section~\ref{0}.

The rest of the paper is structured as follows: Section~\ref{0} presents previous work from the software engineering education domain and beyond. Section~\ref{1} provides an overview of the course, both in general and in respect to the changes made when the course was flipped. Section~\ref{2} details the statistical analysis of both the exam grades and the course evaluations are presented in detail. Results are discussed in Section~\ref{3}, and in Section~\ref{5} conclusions are drawn and future work is considered.

\section{Previous Work}\label{0}
Numerous studies in educational research and in engineering education report on the effectiveness of active learning \cite{freemanpnas,prince2004does}, providing clear evidence due to their being large secondary studies with huge sample sizes. The conclusion of Freeman et al.\ \cite{freemanpnas} that active learning increases student performance, is based on 225 primary studies. The evidence of the whole concept of the flipped classroom is less clear. In a secondary study, Bishop et al.\ \cite{bishop2013flipped} conclude that most existing studies up until 2013 look at student perceptions and only include single-group study designs. The results of student perceptions of the flipped classroom were found to be somewhat mixed, but positive overall. The authors also reported that students tend to prefer in-person lectures over videos, but that they also preferred active learning to traditional lecturing. Bishop et al.\ \cite{bishop2013flipped} also concluded, in 2013, that there is only anecdotal evidence of improvement in student learning in the flipped context, and recommended future work to study the effects more objectively and by using experimental or quasi-experimental designs.

In more recent studies, Kerr \cite{kerr2015flipped} conducted a short survey of the research and found an increase in studies in this topic in engineering education in 2015. The studies show high student satisfaction and increased performance using the flipped classroom methodology in engineering education; the research methods included discourse analysis, quasi–experimental designs, and mixed methods. However, she concludes that a lot of studies do not include statistical analysis of the data nor do they have enough details about the context of the instruction.

In a very recent 2018 review, Karabulut-Ilgu et al.\ \cite{karabulut2018systematic} analyzed papers up until May 2015 and conclude that research on the flipped classroom in engineering education focuses more on documenting the design, but these were only preliminary findings. The authors call for more studies with sound theoretical frameworks and evaluation methods to establish the methodology in engineering education.

The software engineering context has also seen an increase in studies in the last couple of years. Paez et al.\ \cite{paez2017flipped} obtained positive results from flipping a software engineering course, but had a small sample and no control group. Lin \cite{lin2019impacts} conducted an experimental study published in 2019 on using the flipped classroom approach to software engineering students, and concluded that there was an increase in a diversity of aspects, such as an improvement in the students' learning achievement, learning motivation, learning attitude, and problem-solving ability. However, this study was conducted in one course offering as quasi-experiment. Another recent study by Erdogmus et al.\ \cite{erdogmus2017flipping}, who shared their experiences of flipping a software engineering course (also a single course offering) in 2014. They summarize some of their challenges and note that they underestimated the number of teaching assistants needed for a flipped approach. They also, throughout the course, offered students more opportunities to share their learning with peers. For the study reported here, three flipped classroom pedagogical experts were recruited, four student teaching assistants were hired, and student outcomes were monitored across four years.

Previous work on this topic motivated this longitudinal study of the effects of flipping the classroom with rigorous statistical methods for software engineering students by measuring changes in both students' grades and course evaluation.

\section{Course Implementation}\label{1}

The effect of flipping the Empirical Software Engineering (ESE) course (compulsory in the first year of the Software Engineering Master's Program) was evaluated by comparing four fall semester offerings, of which the first (given in 2014) was a traditional lecturing course and the following three were flipped courses (2015-2017). The teaching team was constant for the first three years, except for teaching assistants who corrected assignments and participated in the in-class activities when the course was flipped. In the last year (2017), to test if the effect on grades was kept even if the teachers changed, the flipped course was taught by two new staff, who were Ph.D.\ students; the previous teacher, for the most part, only participated in a few classes and acted more as a support function to the new teachers.

In addition, parts of the course were run at the University of Zambia to evaluate how teaching a flipped version of empirical software engineering differs in a different culture. However, grades were not collected from that short pilot course, only feedback in form of a course evaluation questionnaire. For the course given in Sweden, both grades and course evaluation questionnaires were collected for all the four years the course was given.

The ESE course is worth 7.5 credits, equivalent to 20 hours expected work per week for the students, and was given in November and December each year. It teaches the basics of empirical software engineering, with a focus on applied statistics for the commonly used methods in the software engineering research field. The specific topics taught in the course and the student learning objectives can be found on the Chalmers University course web page\footnote{\url {https://student.portal.chalmers.se/en/chalmersstudies/courseinformation/Pages/SearchCourse.aspx?course_id=28866&parsergrp=3}}.

The course was organized around 14 double lectures (two 45-minute sessions with a 15-minute break), and three laboratory assignments of the same length in which students, in groups of about four, used statistical software to solve an assignment. The first lab comprised using statistical software on real software engineering data (or data taken from the course book on experimentation in software engineering) to output different types of descriptive statistics and to decide what they mean in relation to the data collected. In the second laboratory assignment, students were also given data sets from real examples, but were instead instructed to use inferential statistics and interpret the results. The third laboratory assignment, the Paper Helicopter Experiment\footnote{\url {http://www.paperhelicopterexperiment.com/}}, was not in software engineering, but served to give students hands-on experience on factorial experiments through more active learning. For all three labs, each group handed in compulsory lab reports, which were graded Pass or Fail. Since an overwhelming majority of the student groups pass the laboratory assignments after a couple of iterations with the teachers, these grades were not considered when assessing the effects of the flipped classroom approach.

\subsection{Traditional Course}
In first year of this study (2014), the entire course was given using classical lecturing for 50 students, who served as a control group. Every lecture was taught using PowerPoint slides while the students took notes. Occasionally, the teacher drew on the blackboard in response to student questions. The lectures were based on two textbooks; students were given a schedule of lectures and their correspondence to the book chapters. For some lectures, a set of research articles was also included and made available on the course web page, which only included the syllabus, schedule, and reading material. The three labs were the same for all four years and began with a brief introduction by the teacher that the lab instructions were online. After this short introduction the student groups worked independently, and the teacher and the teaching assistants (former Master's students who had taken the course the previous year) walked around the two classrooms ready to answer questions and offer help.

\subsection{Flipped Course}
The second, third, and fourth years served as the experimental group, with some variation in the fourth year. The flipped version of the course was organized around 11 active learning double classes (again two 45-minute sessions with a 15-minute break). Five of these were completely flipped, meaning that the entire 90 minutes were devoted to material that the students were supposed to have gone through beforehand.

\subsubsection{Pre-Class Activities}
The material was on an online platform and consisted of video lectures of around 10 to 20 minutes each, packaged in the form of slides with text wrapped around the embedded videos. Two video lectures were recorded by the teacher and the rest were taken from YouTube, where good quality videos on the topic were available, since introducing basic applied statistics (like in the first class) is a general subject in most disciplines. It has been reported that good quality videos are hard to find in some disciplines \cite{herreid2013case}. The online material also had four larger quizzes, one after each online component connected to an active lecture. A quiz consisted of around 30 multiple choice questions each with four alternative answers. For example: ``Helena is a software engineer at a car manufacturer. She is developing a car software component for a self-driving car that predicts sensor failure. She thinks that she needs to take latency in data transfer, signal strength, and noise (error in data) into account in her model. She plans to test the car and collect failure data in number of failures and the other data on two levels: Latency (20 and 40), Signal strength (10 and 80), Noise (3 and 5). How many factors does Helena have in her experiment? a) 8, b) 3, c) 4, or d) 2.” and ``If Helena wanted to reduce on the number of unitary experiments she has to conduct because she wants to save time and money, which of the following approaches should she use? a) Two-Way ANOVA, b) One-Way ANOVA, c) Fractional Factorial Design, or d) Full Factorial Design.''

After each video the students were also asked to solve multiple choice questions on the videos and given the opportunity to provide open feedback about what was difficult and what they wanted the in-class discussions to focus on. Three of the active learning classes were labs and were the same as before the course was flipped, and three of the active learning classes were revision lectures in which students were given the opportunity to decide what should be discussed and further explained in class.


\subsubsection{In-Class Activities}
After having put all the lecture material online, teachers need to plan the now-empty lectures. It can be challenging to plan the active lectures, since they need to be planned the night before or the same day as the lecture. Students tend to delay looking at videos and providing feedback, and it takes a lot of teacher team resources to plan effectively often only hours before the actual lecture. However, prompt student feedback is essential, and the beginning of each lecture typically consisted of teachers' comments and explanations of difficulties reported by students after watching the videos. In the first two years of flipping the classroom the pedagogical experts helped to create the online components and plan in-class activities, as well as participating during class and providing feedback.

The active lectures had a mixture of: (1) An introduction with students having a five- to ten-minute discussion in pairs on the corresponding online component---what was it about? was it difficult? etc.; (2) A discussion with the whole class, or in groups or pairs, on the online components, which often included an open question that asked students to write about what they had just seen. The discussion focused on which descriptions were troublesome and how they could be improved and made more accurate. (3) About five minutes of administrative information on labs and lab reports. (4) A group discussion about an online video, which students were asked to consider from new perspectives introduced by the teachers. (5) The teachers showed a provocative but accurate statement about an aspect of the subject and the students did think-pair-share \cite{kothiyal2013effect}. (6) An example was worked by hand by the teachers on the blackboard, but at each step the teachers wrote three similar but different formulas, and the students used clickers on their electronic devices to vote for an option. The distribution of the answers was used to pair students with someone of a different view, to then discuss why they voted differently. The whole example calculation proceeded in this manner. (7) An example of a result was shown on a slide and the students did think-pair-share in relation to possible explanations to the result. (8) In response to statements about how to investigate some phenomenon, students were asked to design a study for that context.

\emph{One such example was students being introduced to the fact that software engineering research should include more experiments. The teacher then asked students to design an experiment to measure ``lecture quality'' for the benefit of the teacher and the university. Students worked in groups using the experiment planning protocol in the textbook, listing issues such as definitions of concepts, obtaining accurate measurements, confounding factors, replication, and so on. Students discussed the need to be careful, and skeptical, in experimentation in complex adaptive systems (i.e., software development organizations)}. (9) Some examples were more technical and the student were asked to provide elements of an experiment in class, for instance \emph{an experiment investigating latency in relation to different server programming languages}.

\subsubsection{Regular Lectures in Parallel}
In parallel, six regular lectures were held, independent of other classes both in relation to content and in being taught by a separate teacher who was not involved in the flipped classroom project. Students are known to differ between years in both motivation and prior knowledge of the topic, so to allow within-year comparison some topics were kept in the classical lecturing format. On average, two active lectures and one traditional lecture were given each week. In the fourth year (2017) the online platform and the main teaching team were changed to see if the improvement remained.

The mini-version of the course given in Zambia had only three flipped classes and two labs, since it only lasted for two weeks. However, the course during the rest of the semester was given as regular lectures, so a comparison between a flipped and a regular part of the course was still possible.

\section{Evaluation}\label{2}
The evaluation of this pedagogical experiment had two parts: (1) analysis of student grades across the four years, and (2) extra questions comparing the flipped and regular parts of the course, added to the student course evaluation questionnaires completed by students after every course.

\subsection{Comparison of the Exam Grades}
The course is given at two universities in Gothenburg, Sweden, simultaneously. Chalmers University of Technology awards grades Fail, 3, 4, and 5 (with 5 being the highest grade possible) and the University of Gothenburg (GU) awards grades Fail, Pass, and Pass with Distinction. To compare exam results between years, the corresponding ``Chalmers grade'' was used for all students. The Kruskal-Wallis statistical test, based on ranks, was used, since this allowed the inclusion of the Fail grades that incorporate all the exam results below 3. Grades given to students were based on the intervals: Maximum points: 35, Grade 3: 17--24, Grade 4: 25--30, Grade 5: 31--35. Students' group lab reports had to earn a Pass grade for a student to pass the entire course (the labs were graded Pass\slash Fail).

The typical exam included: (1) A multiple choice question, for instance about the parametric assumption of data or heteroscedasticity. (2) At least one open question about an important statistical concept, such as the relationship between types of statistical error in hypothesis testing or distinguishing between a sample and a population. (3) At least two questions on other research aspects such as sampling, supervised on unsupervised survey research, or ethics. (4) At least one question about assumptions for different statistical tests (like the $t$ test or linear regression). (5) A question about how to design experiments in the software engineering context.
(6) A question in relation to interpreting, or setting up, hypotheses or what conclusions that can be drawn from statistical software output. (7) One larger calculation of a research question that can be solved through the use of analysis of variance (ANOVA). A sample question is: \emph{``A software development company wants to test three different software testing techniques (exploratory testing, unit testing, and integration testing) to see if it will affect amounts of post-release defects. The company has ten software testers and they want you (the experimenter) to block the effect of the different levels of experience among these testers. In other words, they are only interested in differences between the testing techniques. In your design, each tester will test the same part of the system by applying only one of the testing techniques at the time, however, all testers will apply all techniques. The following sum of squares were obtained: the block (testers) = 280, the treatment (testing technique) = 90. The total sum of squares was 500. a) State the null and alternative hypotheses. (1p), b) Set up an ANOVA table and analyze the effect of the treatment (alpha = 0.05). (2p), c) Calculate the effect size where relevant. What is an effect size? (1p), d) Interpret the results in words. (2p), e) What did they gain by using a block design? Based on the results, do you think they did well in their decision to use a block design? (1p), f) What would the results have been if you would not have used blocking? Set up a new ANOVA table and interpret the results. (2p), g) Could you have carried out multiple t-tests instead of the ANOVA? (1p)}.

As an example, the final exam question was graded using the protocol: a) 0.5 point for the null hypothesis and 0.5 point for the alternative one. b) 1 point for $df$ and $MS$ and 1 point for $F$ values. c) 0.5 point for calculating the effect size. and 0.5 point for the definition. d) 2 points for the correct interpretation. e) 0.5 point for each of the two questions. f) 1 point for ANOVA and 1 point for correct interpretation. g) 1 point for stating alpha inflation.

The distribution of grades from each year is shown in Table~\ref{grades}. The author wrote the 2014 exam, and in following years the TAs or the new teachers wrote those exams, at the same level of difficulty to remove bias in both directions, not making the exam easier nor harder than in 2014.

\begin{table*}
\renewcommand{\arraystretch}{1.5}
\caption{Distribution of Grades 2014--2017.}
\label{grades}
\centering
\begin{tabular}{p{1cm}p{3cm}p{3cm}p{3cm}p{3cm}p{2cm}}
{\bf Grade}  & {\bf2014 (non-flipped)}  \# of students (percentage) & {\bf2015} \# of students (percentage)& {\bf2016} \# of students (percentage) & {\bf2017} \# of students (percentage)& {\bf Grand total}\\
\hline
F & 10 (20\%) & 9 (17\%) & 7 (12\%)& 14 (25\%)&40\\
\hline
3 & 28 (56\%)  & 31 (60\%) & 14 (23\%)& 8 (14\%)&81\\
\hline
4 & 10 (20\%)  &11 (21\%)& 22 (37\%)& 30 (53\%)&73\\
\hline
5 & 2 (4\%)  & 1 (2\%) & 17 (28\%) & 5 (9\%)&25 \\
\hline
Year Total  & 50 (100\%) &52 (100\%)  &60 (100\%) &57 (100\%) &219\\
\end{tabular}
\end{table*}

As mentioned, the non-parametric Independent-Samples Kruskal-Wallis Test was used, which is based on ranks, and is an overall test of any differences across all years. The results were: Test Statistic $= 25.812$, $p$(2-sided) $= 0.000$, $N=219$, meaning that there are differences between years, but to know between which years, post-hoc pairwise comparison tests are needed. The mean rank for each year was 89.34 (2014), 89.24 (2015), 137.88 (2016), and 117.72 (2017) respectively. There were significant pairwise comparisons ($p<0.001$) for two comparisons, namely between 2015 and 2016, and 2014 and 2016. The non-significant comparisons were between 2014 and 2015 ($p=1.000$), 2015 and 2017 ($p=0.082$), 2014 and 2017 ($p=0.091$), and 2016 and 2017 ($p=0.424$).

The result means that there was no significant improvement in grades the first year of the flipped approach, so no significant change between 2014 and 2015. However, in the third year (2016), the grades were significantly different from the first (2014) and second year (2015). The third year of the \emph{flipped} approach (2017) was not significantly different from any other year, so the positive change from the second year (2016) of the flipped approach did not remain in 2017 when the teaching team changed. This change cannot be explained by the dropout rate, since out of 68 students registered on the course in 2016, only eight dropped out (12\%), while in 2017 of the 94 registered students, 37 dropped out (39\%); there was not an increase in students following the entire course in 2017 resulting in more Fail grades, but possibly the opposite.

\subsection{Comparison of the Student Course Evaluation Questionnaires}

The second part of the evaluation was the student course evaluation questionnaire completed after each course. The number of students answering each survey, and their corresponding response rates were in 2014 ($N= 26$, response rate 50\%), 2015 ($N= 33$, response rate 45\%), 2016 ($N= 22$, response rate 26\%), 2017 ($N= 34$, response rate 36\%), Zambia 2017 ($N= 7$, response rate 23\%). The lower response rates are due to the fact that the university administration distributes the course evaluation questionnaires via email to all students when the course has finished. The first part of the questionnaire consists of questions included in all years of the study, therefore useful to test statistically over time to further investigate the effects of flipping the classroom.

The questions relevant for the effects of flipping the classroom both in relation to the non-flipped version of the course (2014) and investigating the temporal perspective of the effects were:
\begin{enumerate}
\item \emph{What is your overall impression of the course?} Rated from 1 (very poor) to 5 (excellent).
\item \emph{The teaching worked well.} Rated from 1 (disagree completely) to 5 (agree completely).
\item \emph{The course literature (including other course material) supported the learning well.} Rated from 1 (disagree completely) to 5 (agree completely).
\item \emph{The course workload as related to the number of credits was...} Rated from 1 (too low) to 5 (too high).
\end{enumerate}

The same statistical test as for grades above was used and the list below shows which ones of the Independent-Samples Kruskal-Wallis Tests that were significant ($p$<0.05). Which pairwise comparisons were significant between years and their corresponding mean rank in parentheses are also shown, if applicable (all pairwise comparisons are adjusted for multiple tests by using the Bonferroni correction).

\begin{enumerate}
\item The null hypothesis was rejected (Test Statistic = 18.797, $p=0.001$, $N=122$). Sig.\ difference between 2017 (48.75) and 2015 (72.15). 
\item The null hypothesis was rejected (Test Statistic = 15.288, $p=0.004$, $N=122$). Sig.\ difference between 2014 (50.33) and 2016 (81.00), and between 2017 (51.21) and 2016 (81.00).
\item The null hypothesis was \emph{not} rejected (Test Statistic = 5.909, $p=0.206$, $N=120$).
\item The null hypothesis was \emph{not} rejected (Test Statistic = 3.040, $p=0.551$, $N=119$).
\end{enumerate}

In summary, the only significant difference of the overall impression of the course was between the first year of flipped (2015) and the third year of flipped (2017), meaning that the students assessed their overall impression higher in 2015 than in 2017. The only parts that were changed were the main teachers and a new platform for the online components. Perhaps the answer lies in the implementation and motivation of the flipped approach.

The same introduction was given in both versions of the course on why the classroom was flipped, but in the open comments section of the survey there were more comments from students not seeing the point of the flipped classroom in 2017 than in 2015. Maybe in 2015 the students were more tolerant, having been clearly told that this was the first time parts of the course would be flipped. But in both years, some students also commented that they really like the flipped approach. Another explanation could of course be that the new teachers in 2017 were less liked by the students, possibly because the new teaching team was new to the flipped approach. This difference did not hold for a comparison between 2016 (when the grade were significantly higher) and 2017. There were no changes in the planned activities in the classroom, however, the teacher during 2015 and 2016 did better in facilitating the students' discussions in class, going from only asking questions of the whole class to physically walking around and listening in to the discussions held in small peer-groups and directly asking student groups about sharing aspects that the teacher had heard. One example would be: ``I heard you [pointing at a specific group] raised the very interesting aspect [X], would you mind sharing that aspect with the whole class?'' Such encouragement significantly increased student participation in class, probably more than in 2017. As an aside, an auditorium is not an optimal lecture hall to use for the flipped classroom because student sit in rows and cannot move around easily. No better-designed lecture hall existed at the university campus. In all the years, some students stated that it would have been easier to have all their courses flipped, since they need to act very differently depending on the teaching methodology used.

Students thought that the teaching worked better in 2016 (the second version of the flipped course) than the non-flipped version in 2014 and the third year of flipped in 2017. In 2017 and 2015 the students, though, both praised and criticized the flipped approach about equally. In 2014 the students complained a lot about how inappropriate it was to use lecturing with slides for teaching the subject. Something essential seems to have happened in 2016 since the grades changed significantly from 2015. The simple explanation for this change is that 2016 was the second year  of teaching the flipped course for the same teacher, which apparently meant that students learn better using the flipped approach and also earned higher grades. However, this increase in grades did not persist in 2017 when the teaching was graded lower along with the exam results, which might have been due to the new teachers. This highlights the importance of coaching new teachers extensively in the flipped approach since it is much more dependent on teacher facilitation in class and short-term planning before each active lecture. Below are some descriptive statistics on specific questions in relation to a comparison between the flipped and the traditional approach to teaching.

\begin{figure}
\centerline{\includegraphics[scale=0.6]{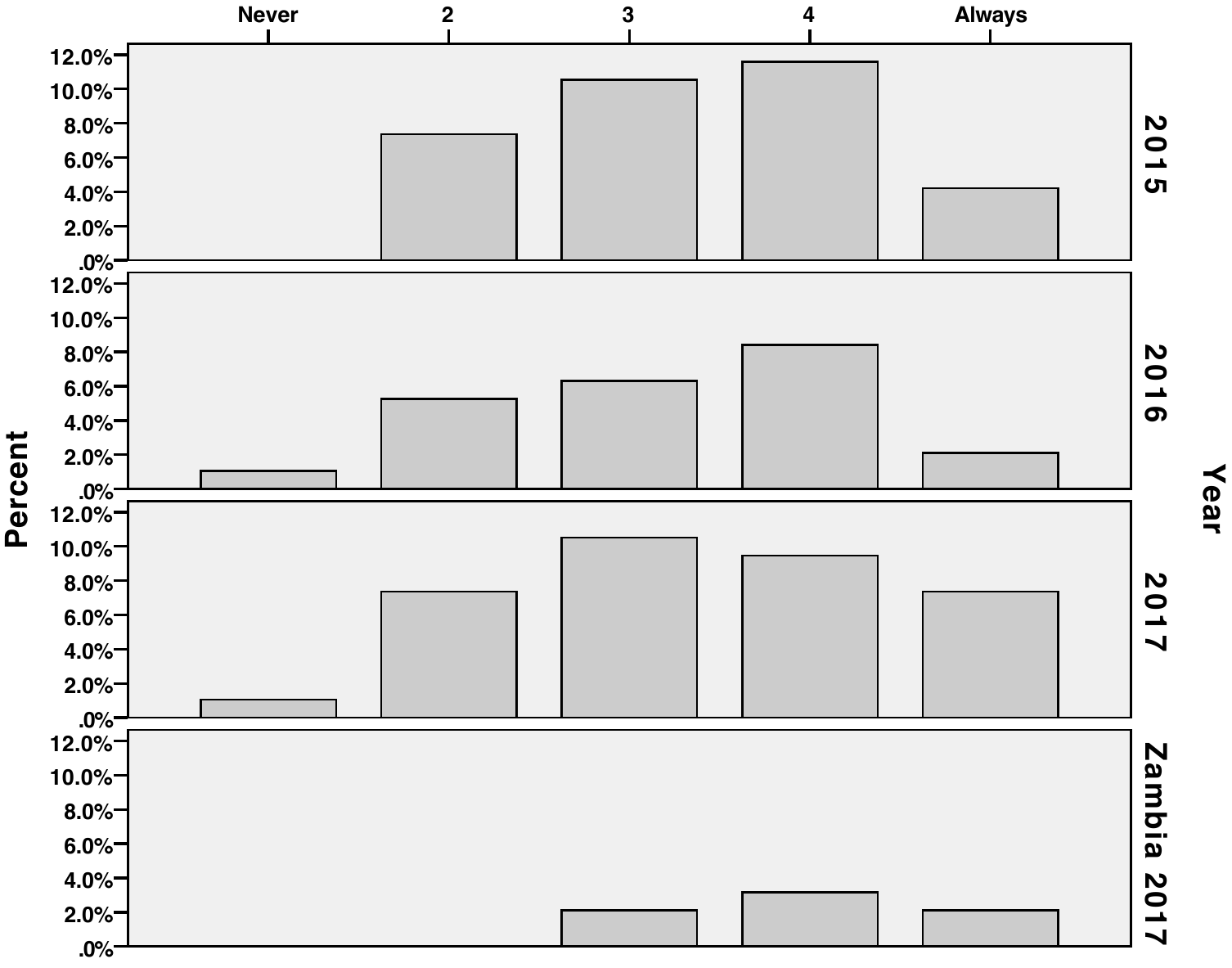}}
\caption{\emph{I was engaged when participating in classroom activities.} Rated from 1 (never) to 5 (always).}
\label{engaged}
\end{figure}

In order to assess how well the flipped classroom approach was implemented across years, the following two questions were included in the questionnaire: 
\begin{enumerate}
\item \emph{I was engaged when participating in classroom activities.} Rated from 1 (never) to 5 (always).
\item \emph{The instructor made meaningful connections between the topics in the pre-recorded lecture and the class activity.} Rated from 1 (disagree completely) to 5 (agree completely).
\end{enumerate}
The descriptive statistics for the first question are shown in Fig.~\ref{engaged} and shows that students considered themselves engaged overall. 
There was no statistically significant difference between the years (i.e.\ we \emph{fail} to reject the null hypothesis, Test Statistic = 2.791, $p=0.425$, $N=95$), which means that it is not possible to explain differences in opinions or grades by the level of engagement by students. For the second question, the null hypothesis was rejected (Test Statistic = 23.053, $p=0.000$, $N=95$), and significant differences between 2017 (33.53) and 2016 (62.61), and between 2017 (33.53) and Zambia 2017 (72.93) were found. These numbers match the student grade differences but not the overall impression of the course by students. This aspect, then, could explain why 2016 was a successful year; making meaningful connections between online material and active lectures could be an important skill for teachers to acquire in order to increase students' academic success. In doing so, the course content will seem stringent and explicitly connecting different aspects that are introduced could increase the number of ``Aha!'' moments for students. The course given in 2016 and the Zambian pilot course were both given by the same teacher. In 2016, the teaching was also rated as working very well (see above). The 2017 version of the course was given by two teachers without any previous knowledge of flipping a classroom.

Finally, the students were also asked to compare the two approaches each year when a part of the course was flipped. They were asked the following questions: 

\begin{enumerate}
\item \emph{Please compare the two approaches in the course (the flipped and the regular lectures).} Rated from 1 (regular is much better) to 5 (flipped is much better).
\item \emph{Please compare the flipped approach to having the same material as regular lectures (i.e.\ the statistics).} Rated from 1 (regular is much better) to 5 (flipped is much better).
\end{enumerate}

\begin{figure}
\centerline{\includegraphics[scale=0.6]{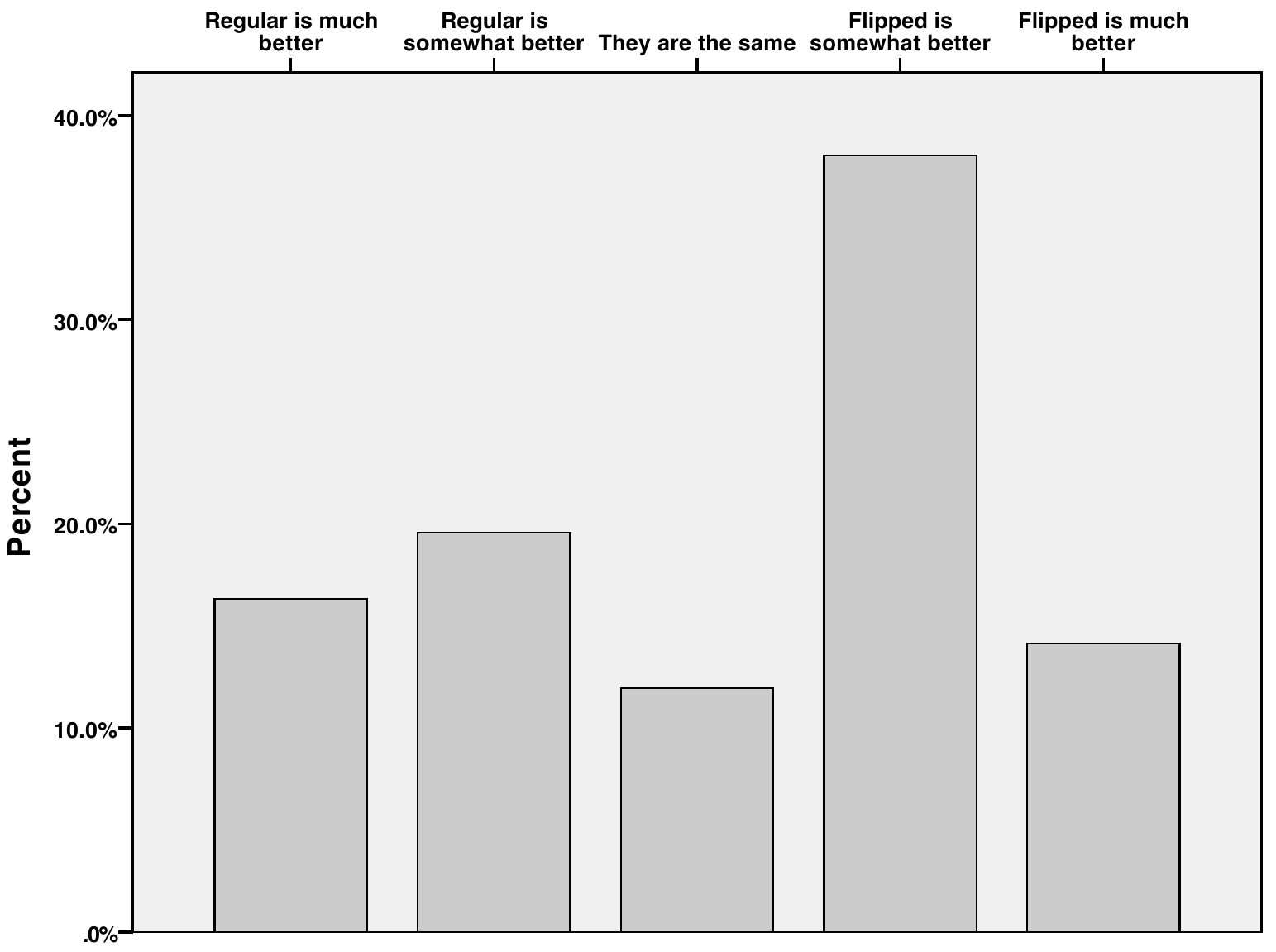}}
\caption{\emph{Please compare the two approaches in the course (the flipped and the regular lectures).} Rated from 1 (regular is much better) to 5 (flipped is much better).}
\label{compare}
\end{figure}

\begin{figure}
\centerline{\includegraphics[scale=0.6]{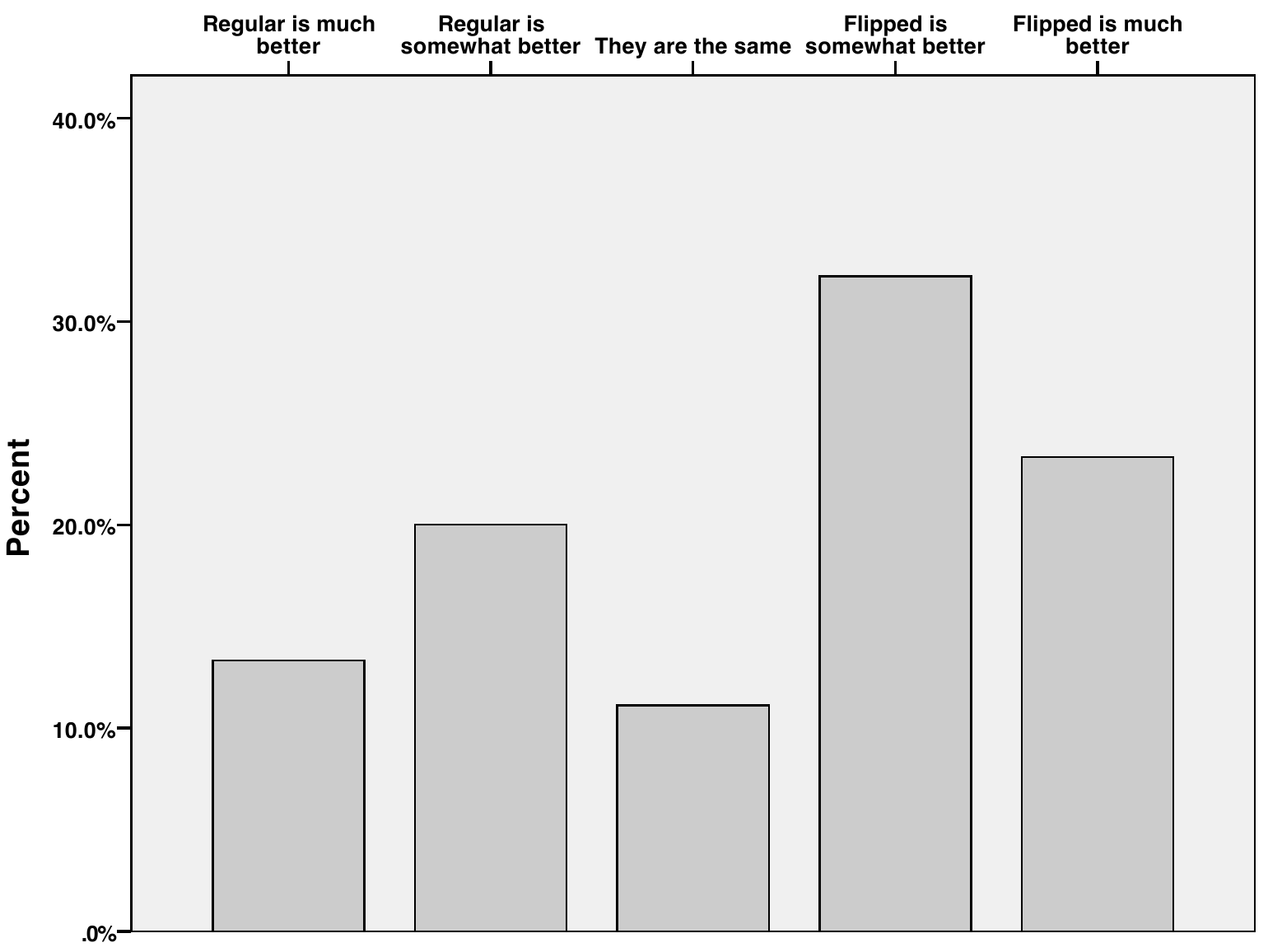}}
\caption{\emph{Please compare the flipped approach to having the same material as regular lectures (i.e.\ the statistics).} Rated from 1 (regular is much better) to 5 (flipped is much better).}
\label{flippedasregular}
\end{figure}

The descriptive statistics for all years are shown in Figs.~\ref{compare} and \ref{flippedasregular}. The figure essentially shows a two-peaked distribution meaning that the students are split between those who think that the flipped approach is better and those who think that a traditional lecturing approach would be better; few students saw them as equal. This means that the students either dislike the flipped approach or like it (a majority tends to like it, Fig.~\ref{compare}), but the classes are split. In the small sample from Zambia 2017, the students all preferred the flipped approach to teaching. When looking at the grades, the changes made towards more active learning through flipping the classroom did result in higher grades; the student preference of pedagogical method does not seem to entirely overlap with actual learning outcomes. Previous studies have shown that student course evaluations are prone to many biases  \cite{spooren2013validity}, as discussed next.

\section{Discussion}\label{3}
Overall, the results from flipping the classroom for an empirical software engineering course were promising. The results from this study are important since few studies have investigated the flipped approach to software engineering education, and also due to the fact that there is a lack of larger longitudinal studies on flipping the classroom in higher education in general \cite{bishop2013flipped}. This implies that the use of the flipped classroom seems promising for teaching more software engineering topics, but only with extensive pedagogical training for teachers. Perhaps the software engineering students are extra susceptible to this type of teaching due to their deep knowledge of IT, but this remains to be further investigated.

The diversity of active components in-class, and students' preparation before class through watching videos online, reflecting on those videos, and doing quizzes on the material, improved grades when the teacher had pedagogical training and experience from flipping a classroom. However, the effect on grades only surfaced in the second year of the flipped approach, which means that teachers implementing a flipped approach should persevere if the effects are not shown immediately. Changing the pedagogical approach is a large change for the teachers as well as for the students, and mastering active learning classes takes time. The evidence on the effects of introducing active learning components, however, is clear \cite{freemanpnas}, and flipping the classroom is a way of buying more time for active learning classes without adding more hours to the course schedule when students and teachers need to co-locate. This study shows that the initial effort of putting good material online and creating good discussion topics in class are not enough. The real challenge starts when teachers have to facilitate active lectures in a way they are not used to.

In this study, the improved effectiveness of the flipped course was confirmed through the significant increase in exam grades, but this effect did not remain when the main teachers of the course were changed together with the online platform used. This could be interpreted as the choice of online learning platform affecting student learning, but for software engineering students, this is somewhat unlikely. Building new teaching skills for this new pedagogical approach, however, and preparing students for the pedagogical change were shown to be important in this current study since the grades were lower after changing the teaching team. The fact that pedagogical experts were hired for the first two years was definitely a key to success in creating both the pre-class and the in-class activities, and bringing in new teachers without external expertise was shown to be difficult. This study also shows that exam results do not fully correspond to the students' perceptions of their learning experience, although the majority of students liked the flipped approach more overall when asked to compare the two rather than only rating their overall impression of the course each year. The results suggest that students' academic success, but maybe not their subjective overall liking of a course, can be enhanced by introducing a flipped classroom and thus more active learning in class.

Also worth discussing is that of the 60-90 students enrolled, often only around 50 finished the course every year. An expected effect of flipping the classroom, could be a lower dropout rate. This was not achieved in flipping the course, but there was an observed increase in how active the students who finished the course were. From previously having around five active students, the estimate for the flipped approach was around 15--20. The high number of dropouts was explained by students as an effect of having two courses with high workload during the same study period. When asked, they said they chose to focus on the other course to a large extent and were planning on passing the ESE course during re-examination periods.

It is important to discuss validity threats to educational research in relation to quasi-experimentation. This study only looked at changes in grades over a four-year period of which the first year comprised of a traditional approach to teaching and the following three years were partially flipped. Even if the exam was not created, nor corrected, by the teacher team when the experiment started, it is very difficult to control aspects like how teachers grade and the level of exam question difficulty. However, the student exam grades are the best available option to investigate effects across many years, but that only holds given the assumption that a written exam is a good measurement of student learning. The exam grades, at least, capture some aspects of student learning even if there are additional aspects to learning not measured during a written exam.

In student course evaluations, the relation to student learning is even more complex. In a meta-study by Spooren et al.\ \cite{spooren2013validity} they conclude that research on the topic is far from having provided clear answer to critical questions and present studies that show that the course evaluations are affected by many different aspects. In relation to the present study the most eminent confounding factors that cause a negative effect in relation to the student course evaluations are:

\begin{itemize}
\item Class attendance \cite{spooren2010credibility} --- there was always a drop in class attendance throughout the course (both flipped and non-flipped); out of 90 registered only around 50 attended the classes.
\item Pre-course interest \cite{olivares2001student} --- in the comments section of the course evaluations many students described learning about research methods and applied statistics as irrelevant for their careers as software engineers.
\item Interest change during the course \cite{olivares2001student} --- the dropout rate could be explained by the course being given in parallel to another course  perceived as being time-consuming.
\item Instructor's tenure \cite{mcpherson2007leveling} --- the classic lecturing given alongside the flipped course was given by a professor, not a Ph.D.\ student, as in the flipped part of the course.
\item Class size \cite{bedard2008class} --- the classes have mostly been very large (from 60 to 90 registered students each year). 
\item Course difficulty \cite{remedios2008liked} --- many students do not have the background in basic statistics needed for the course. 
\item Course discipline \cite{basow2005student} --- the course is a natural science course.
\item Elective vs.\ required courses  \cite{ting2000multilevel} --- the course is compulsory for the Master's program the vast majority of the students are enrolled in. 
\item General education vs.\ specific education \cite{ting2000multilevel} --- the course is broad and comprises many different aspects of empiricism in research. 
\end{itemize}

Therefore, the results of this study should be seen as an indication of a trend. There are, clearly, too many confounding factors to draw wide conclusions, especially from the student course evaluations. The overall trend across the four years of this study is, though, that software engineering students do learn this difficult subject better when more active learning is introduced through flipping the classroom.

\section{Conclusion}\label{5}
This paper reports the effects of flipping the classroom on exam grades and student course evaluations across four years. Through a statistical analysis of data collected between 2014--2017, flipping the classroom was found to increase the students' exam grades, but a clear effect on students' perception of the course was not found. Furthermore, making relevant connections between online material and in-class discussions was found to be a key to student learning, but requires extensive training and a new skill set from teachers. Students did rate flipping the classroom as better overall when asked to compare the two pedagogical approaches, but their overall impression of the course each year gave a less clear result in connection to exam grades. Overall, flipping the classroom increased student learning and it is recommended that the approach be tested in teaching more software engineering topics. These findings are important contributions to software engineering education, but also to educational research in general, since few studies contain such extensive data over more than two years.

In terms of future research, more studies are recommended that use exam grades corrected by other teachers than those teaching the course in order to control for exam variations. Student course evaluations should also be adjusted for bias before being used as a source for teacher evaluation or in research  \cite{mcpherson2007leveling}.

\section*{Acknowledgment}
The course development was funded by Chalmers University of Technology in form of Quality Funding (Dnr C 2014-1712) given by the Swedish Higher Education Authority in 2015.

Teaching parts of the course at the University of Zambia was funded by International Staff Mobility (Dnr E2016\slash 598) in 2017 by the International Centre at the University of Gothenburg. 

The author would like to thank all the people that have been involved in the course during the years of this study: Richard Torkar, Ivar Thorvaldsson, Henrik Marklund, Carl-Adam Hellqvist, Johan Svensson, Mukelabai Mukelabai, Francisco Gomes de Oliveira Neto, Jackson Phiri, David Issa Mattos, Katja Tuma, Mohammad Haghshenas, and Christos Charalampous.

\ifCLASSOPTIONcaptionsoff
  \newpage
\fi

\bibliographystyle{IEEEtran}

\bibliography{references}

\begin{IEEEbiography}{Lucas Gren} is a postdoctoral researcher in software engineering with an Engineering degree in industrial engineering and management. He also has Master's degrees in software engineering, psychology, and business administration management. Lucas performs research within the interdisciplinary field of combining organizational and social psychology with software engineering processes at Chalmers University of Technology and The University of Gothenburg. Lucas' research interests are: collective Intelligence, organizational and social psychology, agile development processes, and statistical methods (all in the context of empirical software engineering).
\end{IEEEbiography}

\end{document}